\documentclass[dvips,12pt]{elsart}
\usepackage{epsf}
\usepackage{subfigure}
\usepackage[dvips]{graphicx}
\newcommand{\eg}{{\it e.g.}}
\newcommand{\ie}{{\it i.e.}}
\newcommand{\vs}{{\it vs.}}
\begin{document}
\begin{frontmatter}
\title{{\small\rm
\rightline{IIT-HEP-00/1}
}
Energy Absorber R\&D\thanksref{talk}}
\thanks[talk]{Presented at the {\sl NuFact'00 Workshop}, Monterey, California, 
May 22--26, 2000.}
\author{Daniel M. Kaplan, Edgar L. Black, Kevin W. Cassel}
\address{Illinois Institute of Technology,
Chicago, IL 60616\\}
\address{{\rm \large and}\\}
\author{Mary Anne Cummings}
\address{Northern Illinois University,
DeKalb, IL 60115}
\begin{abstract}

We describe the program of research and development being undertaken by a
consortium of Illinois university groups to develop liquid-hydrogen energy
absorbers for muon-beam cooling.

\end{abstract}
\end{frontmatter}

\section{Introduction}


Ionization cooling of a muon beam can be accomplished by passing the beam
through energy-absorbing material and accelerating structures, both embedded
within a focusing magnetic lattice.
The rate of change of the normalized transverse emittance with path length is 
then approximately described by~\cite{Neuffer2,Fernow}
\begin{equation}   
\frac{d\epsilon_{n}}{ds} =
-\frac{1}{\beta^2}\frac{dE_\mu}{ds} \frac{\epsilon_n}{E_\mu} + 
\frac{1}{\beta^3}\frac{\beta_\perp(0.014\,{\rm GeV})^2}{2E_\mu m_\mu L_R}\,,
\label{eq:cool}   
\end{equation}   
where $s$ is the path 
length, $E_\mu$ the muon energy, $L_R$ the radiation 
length of the absorber medium, $\beta=v/c$, and $\beta_\perp$ is the 
betatron function of the beam (where the size of the beam 
is given by $\sigma_x = \sigma_y = \sqrt{\epsilon_{n}\beta_\perp/\beta\gamma}$). 

In Eq.~\ref{eq:cool} we see, in addition to the $dE/ds$ transverse cooling
term, a transverse heating term due to multiple Coulomb scattering of the muons
in the absorbers. Since cooling ceases once the heating and cooling terms are
equal, Eq.~\ref{eq:cool} implies an equilibrium emittance, which in principle
(neglecting other limiting effects) would be reached asymptotically were  the
cooling channel continued indefinitely. Since the heating term is proportional
to $\beta_\perp$ and inversely proportional to the radiation length of the
absorber medium, the goal of achieving as small an equilibrium emittance as
possible requires us to locate the absorber only in low-$\beta_\perp$ regions
and to use a medium with the longest possible radiation length, namely
hydrogen. To achieve low $\beta_\perp$, we want the strongest possible focusing
elements. We are thus led to superconducting solenoids filled with liquid
hydrogen as possibly the optimal solution.\footnote{However, lithium lenses
might give an even lower equilibrium emittance than solenoids with liquid
hydrogen, since stronger focusing fields may be feasible with liquid-lithium
lenses than with magnets, and this may overcome the radiation-length advantage
of hydrogen.}

\section{Cooling channel designs}

A variety of solenoidal focusing lattices have been considered for muon-beam
cooling, including the so-called Alternating-Solenoid, FOFO, DFOFO, SFOFO, and
Single- and Double-Flip
designs~\cite{Status-Report,INSTR99,Design-Study,MUCOOL-Notes}.  In
all of these, liquid-hydrogen (LH$_2$) absorbers are used to minimize multiple
scattering. The specifications of the absorbers for some representative
cases are given in Table~\ref{tab:abs}. 

\begin{table}
\begin{center}
\caption{Specifications of typical absorbers (from Ref.~\cite{Design-Study}).}
\label{tab:abs}
\begin{tabular}{l|ccc|l}
\hline\hline
\multicolumn{1}{r|}{Lattice:} & Single-Flip & FOFO1 & FOFO2 &  \\
\hline\hline
Absorber property & & & & unit \\
\hline
length $L$ & 30 & 12.6 & 13.2 & cm \\
density $\rho$ & \multicolumn{3}{c|}{0.0708} & g/cm$^3$ \\
radius $r$ & 20 & 15 & 10 & cm \\
volume $V$ & 38 & 9 & 4 & l \\
power $P^*$ & 160 & 68 & 71 & W \\
temperature $T$ & \multicolumn{3}{c|}{20$^\dagger$} & K \\
pressure $P$ & \multicolumn{3}{c|}{1} & atm \\
boiling point & \multicolumn{3}{c|}{20.2} &  K \\
freezing point & \multicolumn{3}{c|}{13.8} & K \\
window material & \multicolumn{3}{c|}{Al alloy (6061-T6)} & \\
window shape & ellipsoidal & torispherical & ellipsoidal & \\
window thickness $t$ & 300 & 400 & 200 & $\mu$m \\
\hline\hline
\end{tabular}
\end{center}
{\footnotesize $^*$ Assuming $5\times10^{12}$ muons/bunch at 15\,Hz at 200 
MeV/$c$.\\
$^\dagger$ Operation at 1 atm will probably require a somewhat lower operating
temperature to give more ``headroom" with respect to the boiling point; this
will be iterated once the fluid-flow and thermal models have been fully
calculated and verified in bench tests.}
\end{table}

\section{Absorber power handling}

\begin{figure}
  \subfigure[Proposed SLAC E158 target with cooling loop.]
      {\centerline{\rotatebox{90}{\epsfxsize=3.25in\epsffile{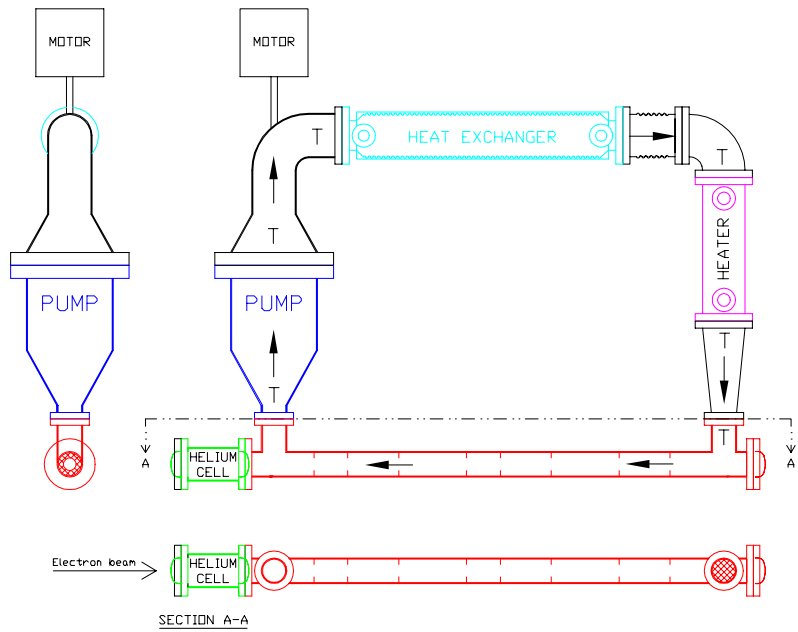}}}}
  \subfigure[SAMPLE target.]
      {\centerline{\rotatebox{90}{\epsfxsize=3.25in\epsffile{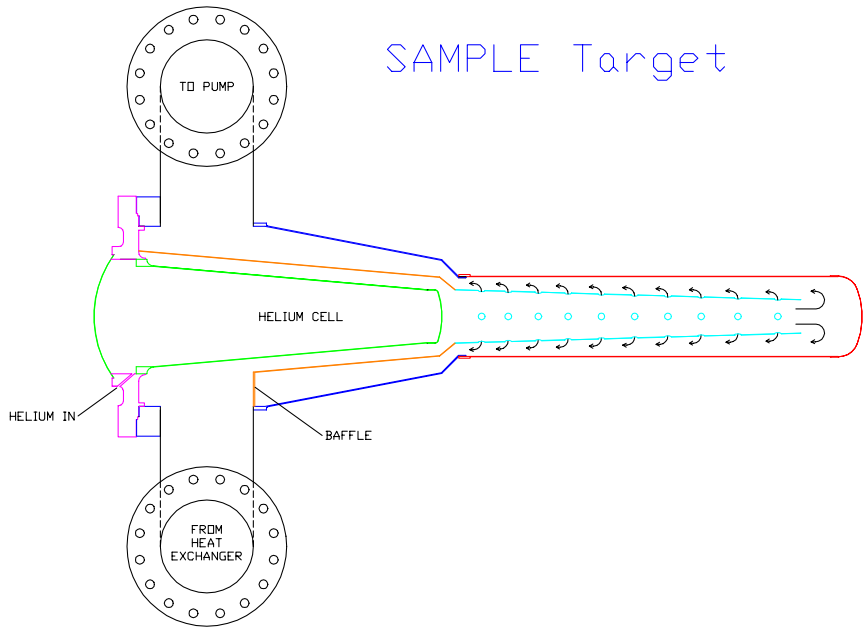}}}}
  \caption{Target designs exmploying an external cooling loop.}
  \label{fig:loop}
\end{figure}

A key problem is handling the heat deposited in the hydrogen by the muon beam.
As shown in Table~\ref{tab:abs}, this can exceed 100 watts per absorber. LH$_2$
targets have been successfully operated in such a heat-deposition regime,
notably the target for the SAMPLE experiment at Bates~\cite{SAMPLE}.  These
high-power target designs have followed the approach developed at
SLAC~\cite{Mark}, in which the liquid is pumped around an external cooling loop
(see Fig.~\ref{fig:loop}). The loop includes a heat exchanger, within which 
the deposited heat is transferred to a supply of cold helium gas, as well as a
heater for fast regulation of temperature. The difficulty in such designs is
assuring flow of the LH$_2$ within the absorber cell that is predominantly
transverse to the beam direction, since any little parcel of LH$_2$ that flowed
along the beam would be quickly heated above the boiling point. In the SAMPLE
target  (Fig.~\ref{fig:loop}b) transverse flow was accomplished using a
specially-designed perforated baffle, and for the proposed SLAC Experiment 158
target (Fig.~\ref{fig:loop}a) the use of asymmetrically-perforated screens is
under investigation~\cite{E158-milestone}. We are developing a design in which
transverse flow within the absorber cell is produced by variously-angled
nozzles placed around the periphery.

\begin{figure}
  \subfigure[Top view.]
      {\centerline{\epsfxsize=5in\epsffile{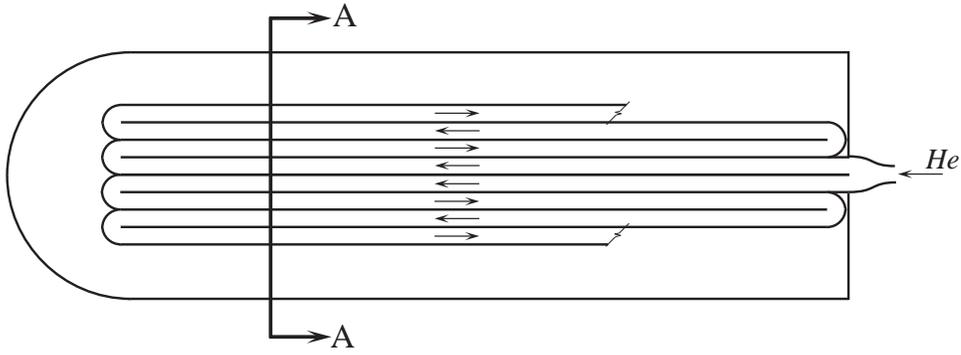}}}
  \subfigure[Section A-A.]
      {\centerline{\epsfxsize=3.25in\epsffile{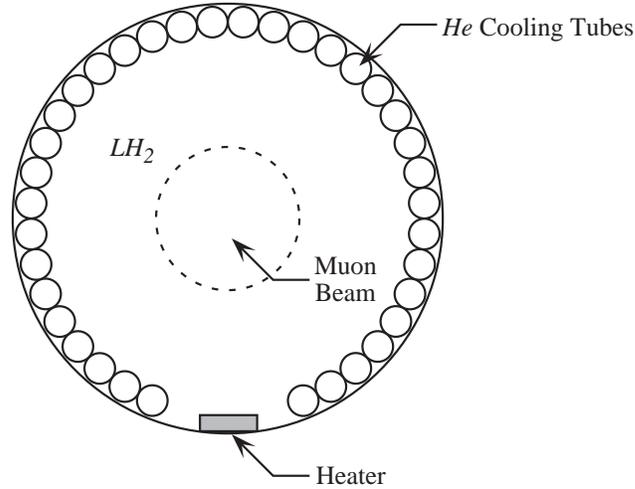}}}
  \caption{Schematic of absorber/heat exchanger design in the convection
  approach.}
  \label{fig:convection}
\end{figure}

In addition to the external-loop approach, we are considering as an alternative
heat exchange internal to the absorber vessel, using driven convection to
provide mixing and transverse flow (Fig.~\ref{fig:convection}). This has the
obvious virtue of fewer moving parts. Moreover, the transverse flow arises
naturally, rather than being imposed by a clever but complicated design. The
key question is whether there will be sufficient convection at the anticipated
power level to give the necessary rate of heat exchange via the cooling tubes
at the periphery of the cell. We are addressing this by a series of
computational-fluid-dynamics (CFD) calculations that numerically solve the
Navier-Stokes equation on a suitable two-dimensional\footnote{Given the
deflection of the absorber windows (discussed below) a three-dimensional
calculation would be preferable but is impractical for any reasonable amount of
computing resources.} grid to evaluate the Nusselt number (the dimensionless
parameter that characterizes the rate of convective heat transport) \vs\ the
Rayleigh number (the dimensionless parameter, proportional to the power
dissipation, that characterizes the degree of turbulence).  Preliminary
indications are that there will be sufficient convection; this will be
investigated further numerically and tested experimentally on the bench.

\section{Absorber windows}

\begin{figure}
\centerline{\epsfysize=4in\epsffile{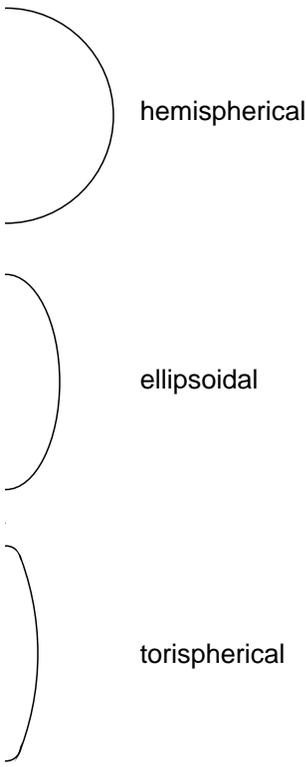}}
\caption{Comparison of ASME window shapes for given absorber diameter.}
\label{fig:windows}
\end{figure}

The
thickness of the absorber windows is a critical parameter. They must be thick
enough to sustain the pressure from the LH$_2$, yet as thin as possible so as to 
minimize
multiple scattering. The window thicknesses in Table~\ref{tab:abs}
have been chosen based on the ASME
standard for pressure vessels~\cite{ASME}. This choice also satisfies the
Fermilab safety code for liquid-hydrogen targets~\cite{FNAL-LH2}. 
As illustrated in Fig.~\ref{fig:windows}, 
ASME~\cite{ASME} specifies three standard window profiles: hemispherical,
ellipsoidal, and torispherical. The minimum thickness required in each case is
\begin{eqnarray} 
{\rm (hemispherical)}\qquad t &=& \frac{0.5PR}{SE-0.1P}, \qquad s = R = 0.5D \\ 
{\rm (ellipsoidal)}  \qquad t &=& \frac{0.5PD}{SE-0.1P}, \qquad s = 0.25D\\ 
{\rm (torispherical)}\qquad t &=& \frac{0.885PD}{SE-0.1P}, \qquad s = 0.169D \,, 
\end{eqnarray} 
where $P$
is the pressure differential, $R$ the radius of curvature (for hemispherical
windows), $D$ the
length of the major axis (for ellipsoidal or torispherical windows), \ie\ the
absorber diameter, 
$S$ the maximum allowable stress, and $E$ the
weld efficiency. In the above equations we give also
$s$, the sagitta of the window at its center, which determines which
window shapes can be used for absorbers of given dimensions: since each absorber
has two windows, the absorber length
$L$ must satisfy $L>2s$. (Note that
for ellipsoidal windows the ASME code considers only the case where the length of the 
major axis is twice that of the minor axis, and for the torispherical case the
radius of curvature of the ``knuckle" is 6\% that of the main cap; see
Fig.~\ref{fig:windows}.) The maximum allowable stress is the smaller of
(ultimate stress)/4 or (yield stress)/1.5~\cite{FNAL-LH2}.  
In practice we find that it is the
ultimate stress that is the limit.

\begin{figure}
\centerline{\scalebox{1.5}{\includegraphics[bb=20 80 300 310,clip]{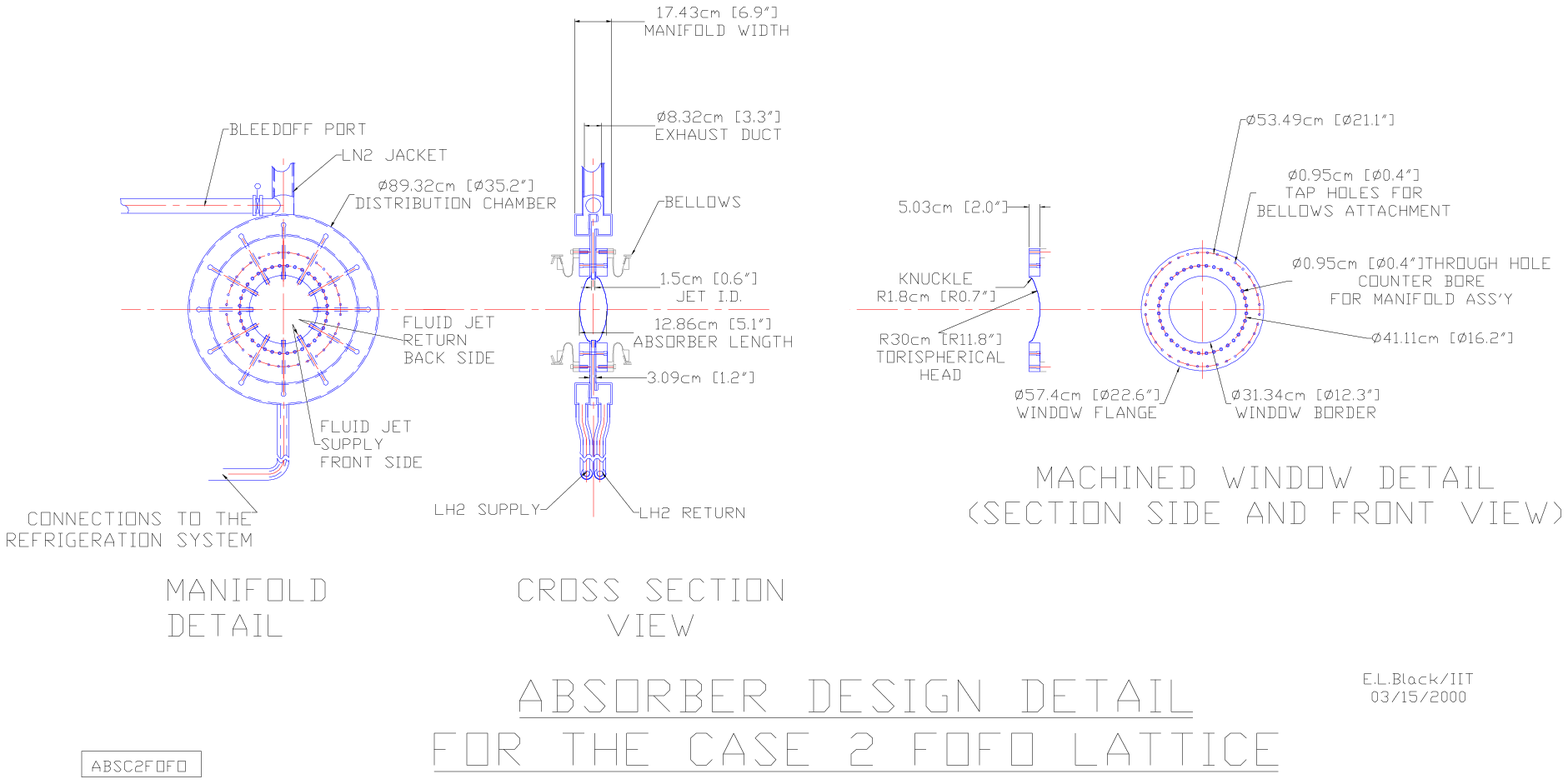}}}
\caption{Mechanical design of FOFO1 absorber (external-cooling-loop version).}
\label{fig:FOFO}
\end{figure}

The hemispherical window shape minimizes the needed thickness. However, a
hemispherical window is practical only for absorbers whose length exceeds their
diameter. Since our absorbers are typically wider in diameter than they are
long, we choose either the ellipsoidal or torispherical window profile. As an
example, we show in Fig.~\ref{fig:FOFO} the mechanical design for the ``FOFO1"
absorber of Table~\ref{tab:abs}. Given the 15\,cm radius of the absorber, both
hemispherical and ellipsoidal windows are ruled out. Torispherical windows with
5.1\,cm sagitta leave just 2.4\,cm for the flange and manifold assembly that
joins the two windows and connects to the hydrogen inlets and outlets. Our
solution is to machine each window with an integral flange out of a single
block of material. The two flanges bolt together in ``clamshell" fashion to
form the manifold. 

In the Neutrino Factory Feasibility Study~\cite{Design-Study} we consider FOFO
and Single-Flip options for the cooling channel. In both designs the cooling
performance is significantly limited by scattering in the absorber windows. To
minimize this effect we propose to operate the absorbers at 1\,atm
pressure.\footnote{Operation of high-power LH$_2$ absorbers at 1\,atm pressure is
not an established technique, 2\,atm having been used in previous high-power
target designs~\cite{SAMPLE,Mark,E158}. There may be safety
concerns that will prevent such operation~\cite{Kilmer-private}; this is a
topic for R\&D.} In the Single-Flip design the absorbers are long enough to
permit the use of 300\,$\mu$m ellipsoidal 
windows.\footnotemark\addtocounter{footnote}{-1}  In the FOFO design two
cooling sections are used having two different absorber sizes. In the first
(``FOFO1") section, 400\,$\mu$m torispherical windows are used as discussed
above.\footnotemark\addtocounter{footnote}{-1} By the second section, the
beam has become small enough to permit a reduction in absorber diameter,
allowing use of 200\,$\mu$m ellipsoidal windows.\footnote{We assumed a weld
efficiency $E=0.9$ in specifying these window thicknesses, however in the
integral-flange approach discussed above, $E=1$, allowing 10\% thinner windows,
or alternatively, operation at 1.1\,atm with the thicknesses given.} 
The reduction in window
thickness  results in a lowering of the equilibrium emittance from 2.6$\pi$ to
2.2$\pi$\,mm$\cdot$rad and a corresponding increase in the cooling rate. 

\begin{figure}
  \subfigure[Uniform thickness.]
      {\centerline{\epsfysize=2.75in\rotatebox{270}{\epsffile{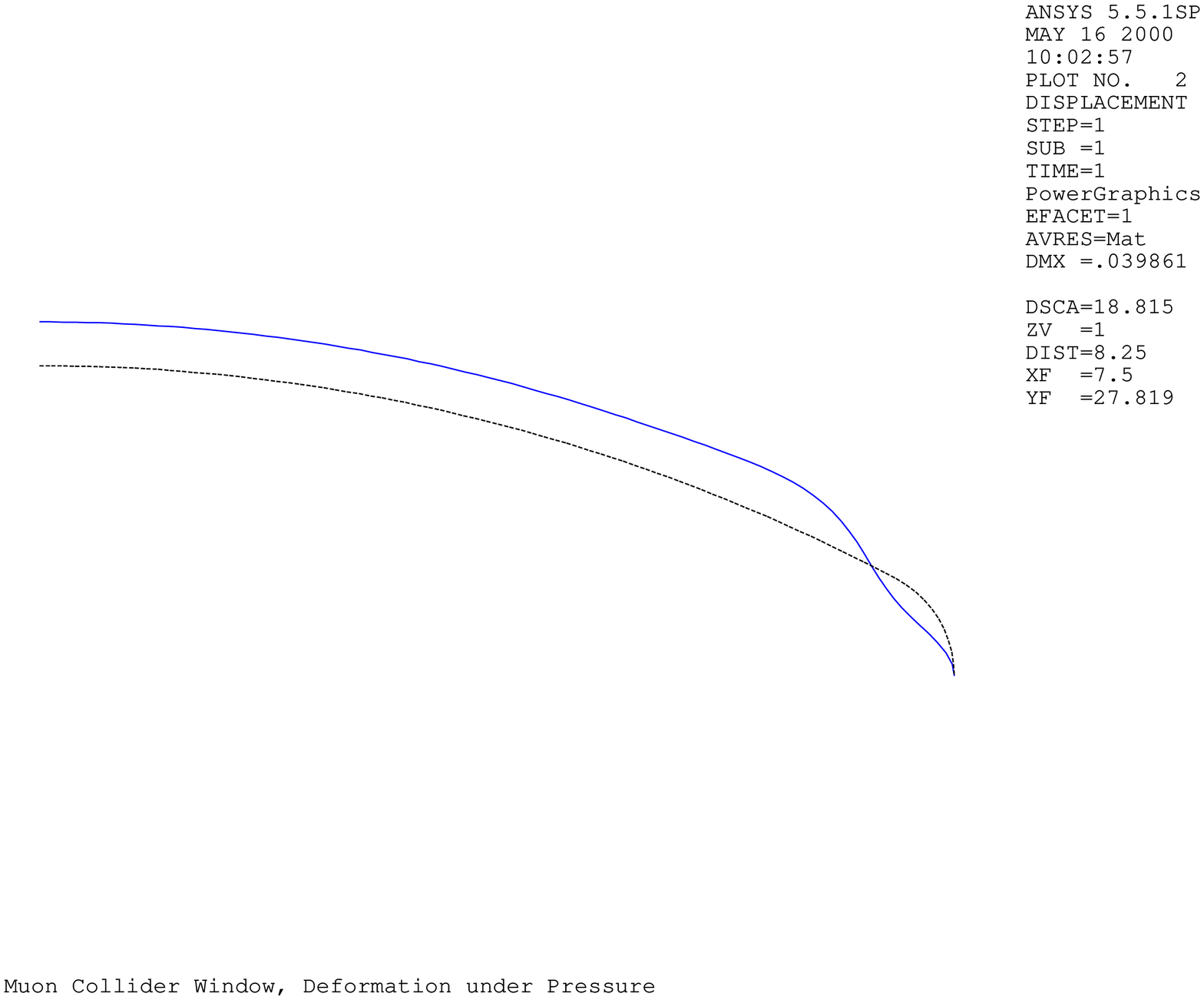}}
      \hspace{-.5in}\epsfysize=3.25in\rotatebox{270}{\epsffile{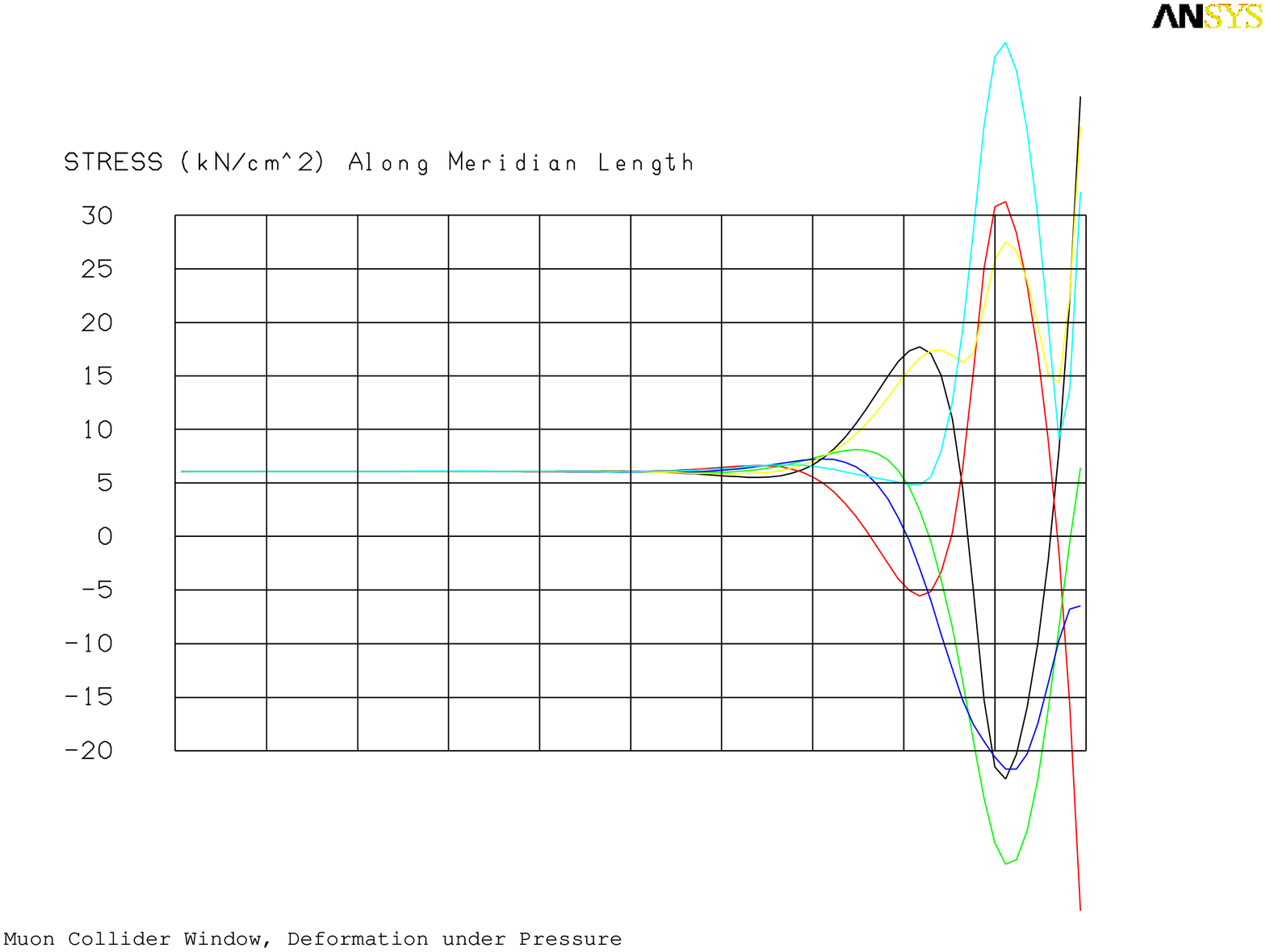}}}}
  \subfigure[Tapered profile.]
      {\centerline{\epsfysize=2.75in\rotatebox{270}{\epsffile{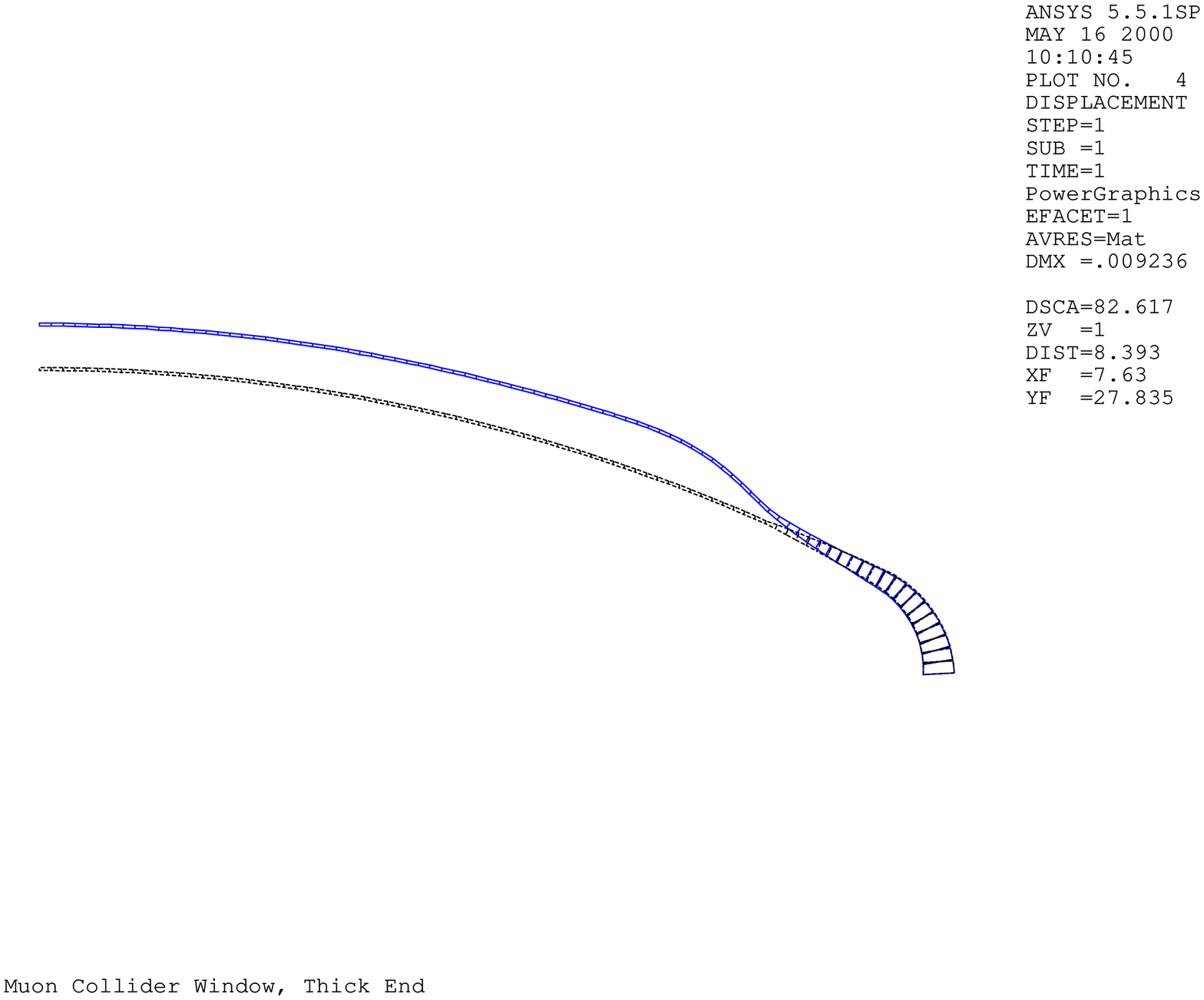}}
      \hspace{-.5in}\epsfysize=3.25in\rotatebox{270}{\epsffile{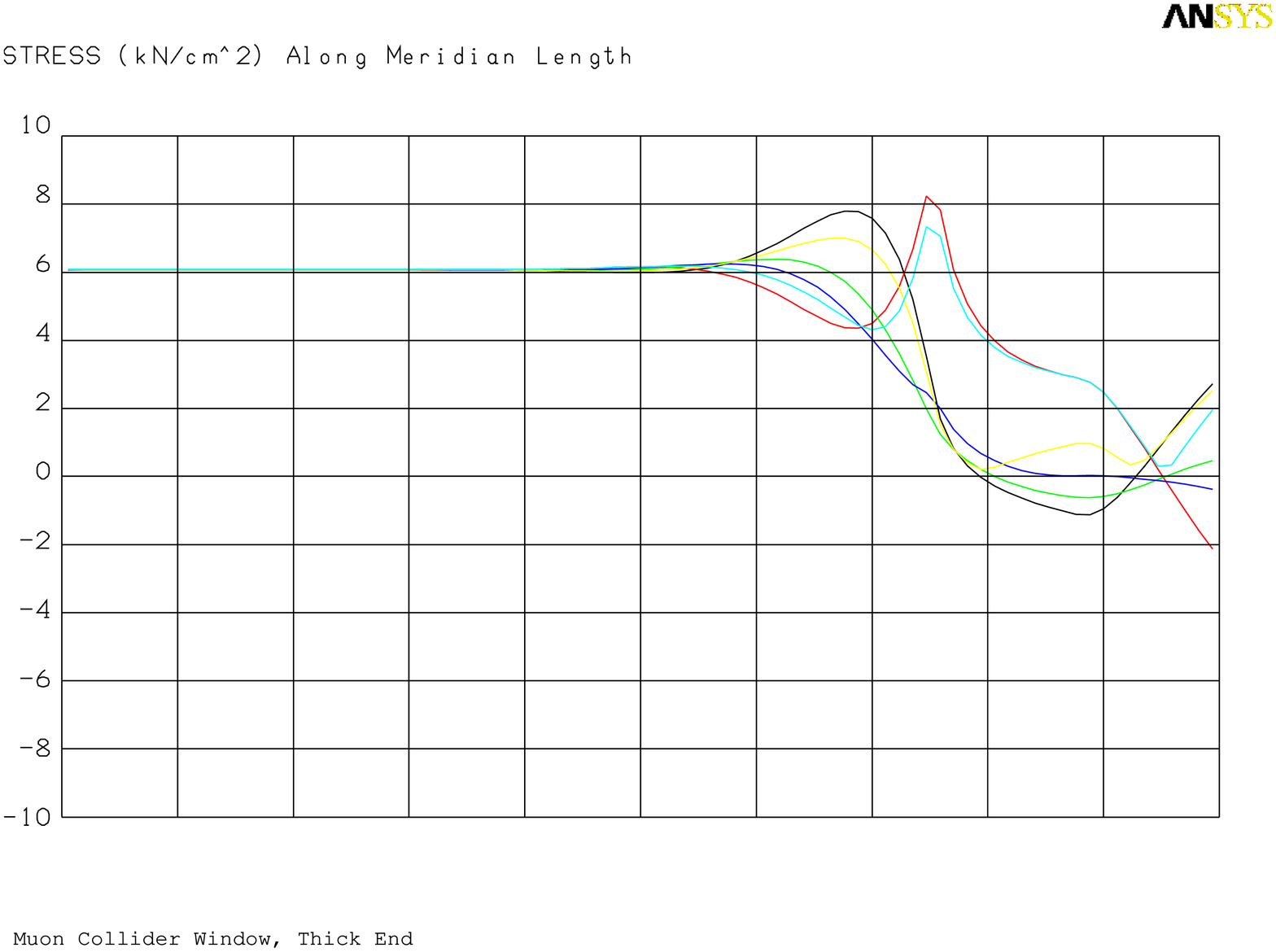}}}}
        \caption{Comparison of deformation and stresses \vs\ position under 
	pressure for a) standard torispherical
  and b) tapered torispherical windows. (Note that the deformation has been
  exaggerated for visibility.)}
  \label{fig:new-window}
\end{figure}

We have also begun to explore the option of customizing the thickness profile
of the window in order to minimize the thickness at the center while maximizing
strength. An ANSYS finite-element calculation has been carried out that shows
that the stresses in a torispherical window are greatest near the edge, in the
region in which the window curvature under pressure exhibits a point of
inflection (Fig.~\ref{fig:new-window}). By thickening the material near the
edge one can reduce the maximum stresses substantially, allowing the material
near the center to be thinner by perhaps as much as a factor of 5. Of course
manufacturability will also impose a limit on how thin the center of a machined 
window can be; we will explore this soon by building and testing prototypes.

\section{R\&D issues}

Beryllium or AlBeMet (a composite of 62\% beryllium/38\% aluminum) could reduce
the impact of the windows on the cooling performance. However, based on the CEA
bubble-chamber accident, beryllium is believed to be incompatible with liquid
hydrogen, and an R\&D program will be required to establish safe design
parameters for these materials. With 40\% greater strength than aluminum and
2.1 times the radiation length, AlBeMet has the potential to lower the total
radiation-length fraction per absorber from ~2.4\% to 1.8\% or less, depending
on the detailed optimization of absorber dimensions. (While beryllium windows
may also be feasible, there may be little additional gain in going beyond
AlBeMet.) Other cooling scenarios (\eg\ SFOFO) use absorbers that are thicker
compared to their diameter. Here effects of windows on cooling performance are
reduced, and aluminum windows may be adequate. Whether R\&D on exotic window
materials is worthwhile may thus depend on which cooling approach prevails. 

In all scenarios
the specific power dissipation in the absorbers is large and represents a
substantial portion of the cryogenic load of the cooling channel. Handling this
heat load is a significant design challenge. An R\&D program is already in
place at IIT to understand the thermal and fluid-flow aspects of maintaining a
constant temperature within the absorber volume despite the large spatial and
temporal variations in power density. This program is beginning with
CFD studies and is planned to proceed to bench tests
and high-power beam tests of absorber prototypes over the next year. 

In some
scenarios (especially those with emittance exchange), lithium hydride (LiH)
absorbers may be called for. Since it is a solid, LiH in principle can be
fabricated in arbitrary shapes. In emittance-exchange channels, dispersion in
the lattice spatially separates muons according to their energies, whereupon
specially shaped absorbers can be used to absorb more energy from muons of
higher energy and less from those of lower energy. However, solid LiH shapes are
not commercially available, and procedures for their fabrication would need to
be developed. Such an effort is challenging since LiH reacts with water,
releasing hydrogen gas and creating an explosion hazard.

\section{Acknowledgements}

We thank M. Boghosian of IIT who carried out the CFD calculations  and Z. Tang
of Fermilab who carried out the ANSYS window calculations. This work was
supported in part by the U.S. Dept.\@ of Energy, the National Science
Foundation, and the IIT Research Institute.

\end{document}